\documentclass[reprint,groupedaddress,nofootinbib,nobibnotes,amsmath,amssymb,aps,prl,twocolumn,showpacs,balancelastpage]{revtex4-1}
\pdfoutput=1

\usepackage{epstopdf} 
 
\usepackage{graphicx}
\usepackage{dcolumn}
\usepackage{bm}
\usepackage[mathlines]{lineno}

\begin{document}


\title{Dynamics of Majority Rule with Differential Latencies}
\author{Alexander Scheidler}
\email[Electronic adress: ]{ascheidler@iridia.ulb.ac.be}
\affiliation{IRIDIA, CoDE, Universit{\'e} Libre de Bruxelles, Brussels,
Belgium}

\date{\today}

\begin{abstract}
We investigate the dynamics of the majority-rule opinion formation model when
voters experience \emph{differential latencies}. With this extension, voters that
just adopted an opinion go into a latent state during which they are excluded
from the opinion formation process. The duration of the latent state depends on
the opinion adopted by the voter. The net result is a bias towards consensus
on the opinion that is associated with the shorter latency. We determine the exit
probability and time to consensus for systems of $N$ voters. Additionally, we
derive an asymptotic characterisation of the time to consensus by
means of a continuum model.
\end{abstract}

\pacs{2.50.Ey, 05.40.-a, 89.20.Ff, 89.75.-k}

\maketitle


\newcommand{\opi}{A}
\newcommand{\opii}{B}

Binary-choice opinion formation models have recently received much attention from
the statistical physics community \cite{Castellano09}. They try to model
consensus formation in populations of interacting voters. Typically, these models
consist of $N$ voters where, at any given time, each voter has one of the two
possible opinions \opi~or \opii. Voter's opinions are influenced by the opinions
of other (neighbouring) voters by means of the repeated application of simple
rules in the population. Prominent and well studied examples of binary-choice
opinion formation models are the Voter model \cite{li85}, the Sznajd model
\cite{citeulike:611482} or the Majority-Rule (MR) model \cite{Galam86}. The
latter was originally proposed to capture the consensus formation in public
debates and has been extensively studied in recent years (see, e.g., \cite{
Mobilia,Krapivsky03, CheRed05, Chen, PhysRevE.71.046123, Lam07,Lambiotte09,
Galam2007366}).

In the MR model the following two steps are repeatedly applied. First, a group of
random voters is selected. Second, the voters in this group adopt the opinion
that is favoured by the majority in the group. The repeated application of these
steps eventually drives the population to consensus, that is, a state in which
all voters have the same opinion. The opinion on which consensus is reached is
determined by the initial fractions of \opi~and \opii~voters. More precisely, the
majority rule amplifies an existing opinion bias: with high probability the
voters end up with the opinion that was initially in the majority.

Lambiotte et. al. \cite{Lambiotte09} extend the MR model with the concept of
latency: after voters have adopted an opinion they go temporarily into a {\em
latent} state in which they cannot be influenced by other voters. However, they
still participate in the opinion formation process and influence other voters.
This extension leads to a rich dynamic behaviour that depends on the
duration of the latent state.

Montes de Oca et. al. \cite{montes10} introduce the concept of {\em differential
latencies} in the MR model. Here the opinion adopted by a voter determines the
duration the voter stays latent. In contrast to Lambiotte et. al. 's model voters
are excluded from the opinion formation if they are latent. As a consequence,
voters that favour the opinion that is associated with the shorter latency
participate more often in the application of the MR. This bias in the opinion
formation process was shown to drive the voters with higher probability to
consensus on the opinion that is associated with the shorter latency. Based on
this finding Montes de Oca et. al. present a decentralized decision making method
for groups of artificial agents. Here the term agent refers to an autonomously
deciding and acting entity like, for example, a robot. Given two possible actions
(opinions) that take different time to execute (latencies) the agents can
collectively find the action which is associated with the faster execution
(shorter latency). For example, it was shown that with the proposed method a
group of robots is able to decide on the shorter of two paths between two
locations without the need to measure travel times. The results presented in
\cite{montes10} are mainly obtained numerically and with normally distributed
latencies.


\begin{figure}[h]
\includegraphics[width=0.47\textwidth]{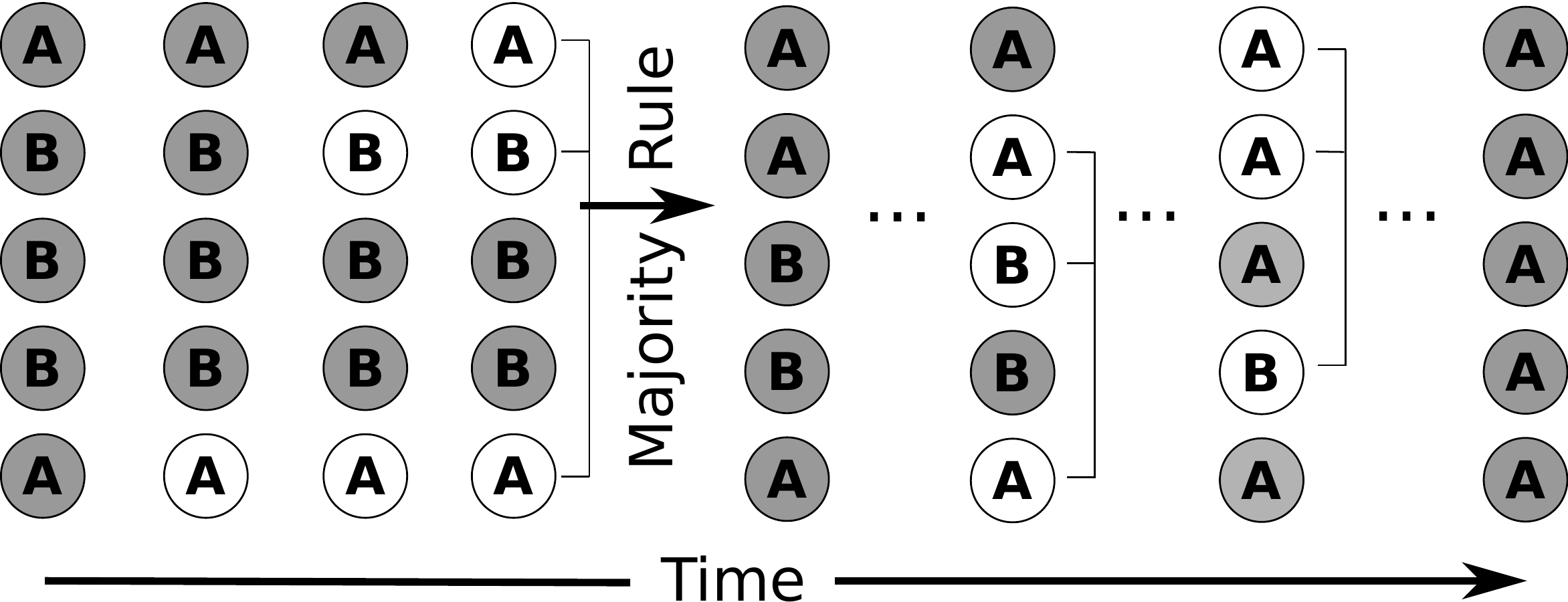}
\caption{Illustration of the update; gray voters are latent, white are
non-latent; as soon as three voters became non-latent the MR is
applied and the voters go back into latent state}
\label{figure:sys}
\end{figure}

The goal of this letter is to study the dynamics of the MR model
with differential latencies analytically  and to support the findings of
\cite{montes10} from a theoretical point of view. To this end, we investigate the following
model (see also Figure \ref{figure:sys}): All voters start latent. The duration a
voter stays latent follows an exponential distribution whose mean depends on the
voter's opinion. Without loss of generality the mean time voters with opinion
\opi~stay latent is 1 and the mean duration of the latent state for voters with
opinion \opii~is $1/\lambda$ with $0 < \lambda \leq 1$. As soon as three voters
have left the latent state the MR is applied and the voters go back into a latent
state. Thus, never more than three voters are non-latent at any given time. Note
that this simplifies the model presented in \cite{montes10}, where an arbitrary
but fixed fraction of all voters stays non-latent.

\begin{center}
{\bf\small Exit Probability}
\end{center}

In the following we estimate the exit probability $E_n$,
that is, the probability that a system of $N$ voters that starts with $n$ voters for opinion
\opi~eventually finds consensus on \opi. Let $n$ be the number of voters that
currently vote for \opi~and $x=n/N$ denote the density of \opi~voters. The
probability $p$ that a voter that leaves the latent state has opinion \opi~is given by
\begin{equation}
p=\frac{x}{x+\lambda(1-x)}
\label{equ:p}
\end{equation}
Note $p$ only dependents on $x$ (we assume $N$ to be large and neglect
that $p$ changes slightly when one or two voters already left the latent state).

$E_n$ obeys the master equation:
\begin{equation}E_n=w_+E_{n+1}+w_-E_{n-1}+w_*E_n\label{equ:master}\end{equation}
with hopping probabilities :
\begin{eqnarray*}
w_+ &=& 3p^2(1-p)\\
w_- &=& 3p(1-p)^2\\
w_* &=& p^3+(1-p)^3
\end{eqnarray*}
We substitute these into (\ref{equ:master}), write $E_{n\pm 1} \rightarrow
E(x\pm\delta x)$ and expand to second order in $\delta x$:
\begin{equation}
 0 = 3p(1-p)(2p-1)\frac{\partial E}{\partial x} + \frac{1}{2} 3p(1-p) \delta x
 \frac{\partial^2 E}{\partial x^2}
\label{equ:expanded}
\end{equation}
Substituting (\ref{equ:p}) into (\ref{equ:expanded}) and letting $\delta x
= \frac{1}{N}$ finally leads to
\begin{equation}
0 = 2N \left (\frac{2x}{x+\lambda(1-x)}-1 \right)\frac{\partial E}{\partial x} +
\frac{\partial^2 E}{\partial x^2}
\label{eqn:de}
\end{equation}
The solution of this equation with respect to the boundary conditions $E(0)=0$
and $E(1)=1$ is
\begin{equation}
E(x)=\frac{I(x)}{I(1)}
\end{equation}
where for the case that $\lambda=1$
\begin{equation}
I(x)=\int_0^x e^{2N(y-y^2)}\mathrm{d}y
\end{equation}
and for $0 < \lambda < 1$
\begin{equation}
I(x)=\int_0^x e^{\frac{2Ny(\lambda+1)}{\lambda-1}}
[y+\lambda(1-y)]^{\frac{4N\lambda}{(1-\lambda)^2}}\mathrm{d}y
\end{equation}

\begin{figure}[h]
\includegraphics[width=0.45\textwidth]{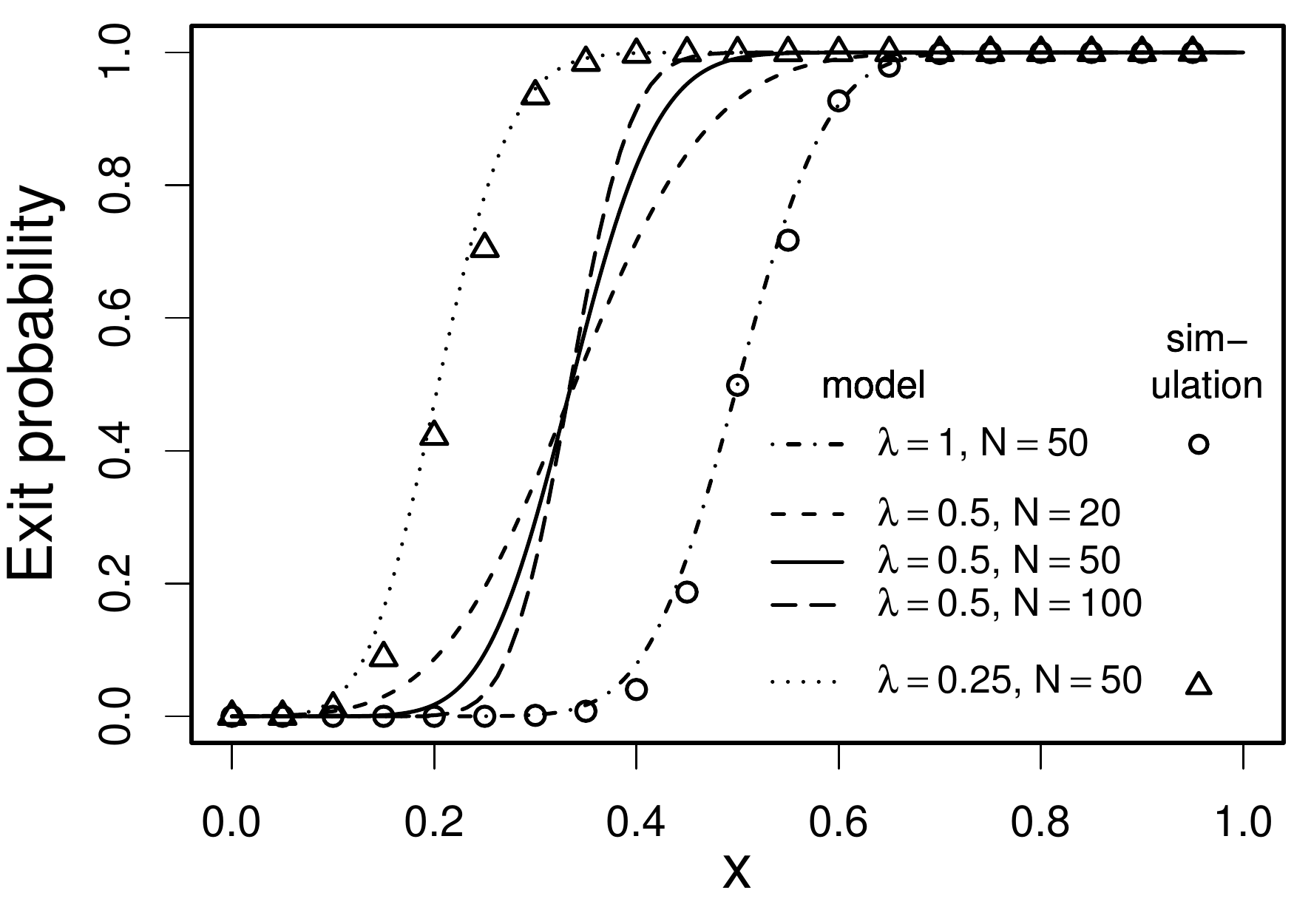}
\caption{Exit probability for 50 voters and different latencies; Comparison of the analytical
model and simulation results}
\label{figure:exit50}
\end{figure}

Figure \ref{figure:exit50} depicts $E(x)$ and results of Monte-Carlo simulations
of 50 voters for latencies $\lambda \in \{1,0.5,0.25\}$ (Note that all presented
simulation results are averaged over 1000 independent runs). Clearly, for
equal latencies ($\lambda=1$) the model is equivalent to the MR model. In this
case the density $x=0.5$ marks the {\em critical (initial) density} of A voters
that determines the consensus state: systems that initially start with $x<0.5$
tend to find consensus on \opii, whereas systems that start with $x>0.5$ find
consensus on \opi~with high probability.

The results for $\lambda \neq 1$ show that differential latencies influence the
exit probability significantly. The more the latencies for the two opinions
differ (the smaller $\lambda$) the more is the critical initial density shifted
towards smaller values. More precisely, the critical density is now given by
$x=\lambda/(1+\lambda)$. This value corresponds to a system state in which the
voters for \opi~and \opii~leave the latent state in the same rates.

In the standard MR model the exit probability converges to a step function for
growing $N$. This is still valid for the MR model with differential latencies
(see the results for $\lambda=0.5$ given in Figure \ref{figure:exit50}).
Clearly, for very large $N$, Formula (\ref{eqn:de}) is mainly determined by the
drift term and only near the critical density the drift term becomes comparable
to the diffusion term.

\begin{center}
{\bf\small Time to Consensus}
\end{center}

%
%
%

The time $T_n$ to reach consensus from a state where $n$ voters have opinion \opi~obeys the master equation

\begin{equation}T_n=\delta t + w_+
T_{n+1}+w_-T_{n-1}+w_*T_n,\label{equ:masterT}\end{equation}

where $\delta t$ denotes the expected time between two applications of the
MR. This value is not constant but depends on the actual fractions
of opinions in the system. To determine $\delta t$ we use the fact that the
minimum of exponentially distributed random variables is exponentially
distributed with parameter equal to the sum of the parameters of the single
distributions. Thus, the expected time between two voters leaving latent state is
distributed exponentially with parameter $\mu=n+\lambda(N-n)$. We can hence
estimate the time between two applications of the MR as $\delta t=3/\mu$ (Note
that this is only valid for large $N$ because it does not take the voters into account that might have already left the latent
state). Inserting $\delta t$ in (\ref{equ:masterT}) and expanding to
second order results in the equation


\begin{equation}
Nx(1-p)(2p-1)\frac{\partial T}{\partial x} + \frac{1}{2}x(1-p)\frac{\partial^2
T}{\partial x^2}=-1\label{eqn:9}.
\end{equation}

\begin{figure}[h]
\centering
\includegraphics[width=0.45\textwidth]{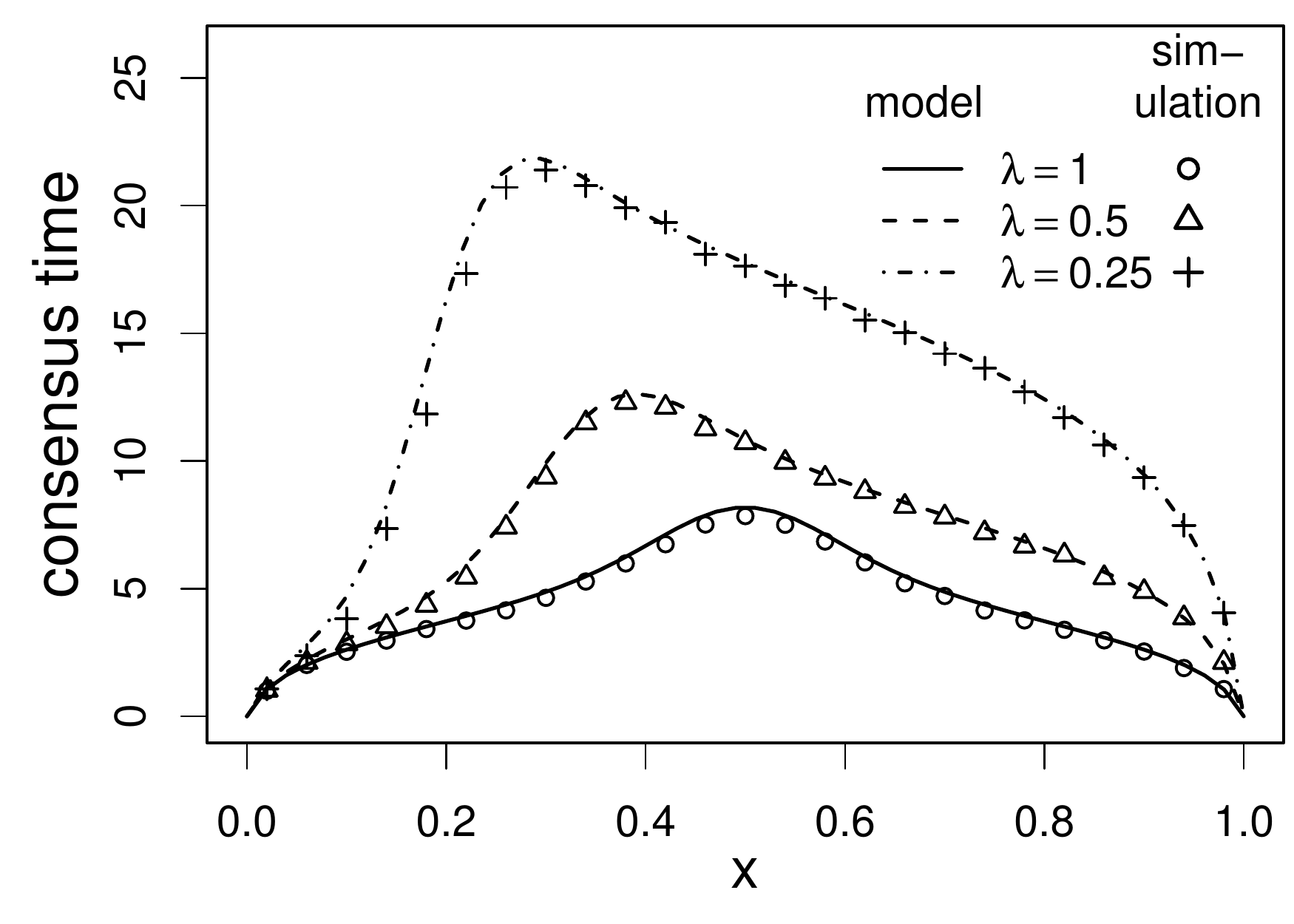}\label{subfigure:ttca}
\caption{Consensus time $T_n^{N}$ versus $n$ (initial number of voters for
\opi) for $N=50$ voters; comparison of the model predictions with results
gained in a Monte-Carlo simulation}
\label{figure:ttc50}
\end{figure}

Figure \ref{figure:ttc50} shows the numerical solution of (\ref{eqn:9}) for
$N=50$ and boundary conditions $T(0)=0$ and $T(1)=0$ in comparison to results
obtained in a Monte Carlo simulation. Without differential latencies ($\lambda
= 1$) the results of the exact solution given in \cite{Krapivsky03} are
resembled. As expected, these values are symmetrical to $x=0.5$ (no initial
bias).

Differential latencies ($\lambda\neq 1$) increase the time to find consensus.
This is because on average less updates per unit of time are applied. Moreover,
caused by the shift of the critical densities the curves are not symmetrical.
Although opinion \opi~is associated with the shorter latency, the time for a
system biased to \opi~takes longer to converge compared to a system equally
biased to \opii~(compare, e.g., $T(0.1)$ and $T(0.9)$). The reason is that the
rate of change mainly depends on the rate voters that are in the minority leave
the latent state. For example, consider a state which is biased to \opi. If
\opii~voters become non-latent there is thus a high probability that they will be
``convinced''. However, this happens only with rate $\lambda$. On the other
hand, if the system is biased to \opii~the rate at which \opi~voters are convinced is $1$
and thus the voters will find consensus faster in this case.

In the following we characterise the asymptotic behaviour of the time to reach
consensus. Formal solutions of Fokker-Planck equations like (\ref{eqn:9}) with
respect to given boundary conditions can be derived (see, e.g.,
\cite{citeulike:1400625}). However, in our case a formal solution is complex and
hard to analyse. We therefore choose to approximate the consensus time by means
of a continuum model. This can easily be done and we show (experimentally)
that this approximation becomes more accurate for large $N$.

In an unit time step the overall fraction of voters that
become non-latent is $x+\lambda(1-x)$. The probability that in a triple of
these voters at least two voters have opinion \opi~is given by $3p^2(1-p) +
p^3$. This leads to the model:
\begin{equation}
\dot{x} = -x + \left [3p^2 - 2p^3 \right]
(x+\lambda(1-x))\label{equation:sdes}
\end{equation}

\begin{figure}[h]
\includegraphics[width=0.45\textwidth]{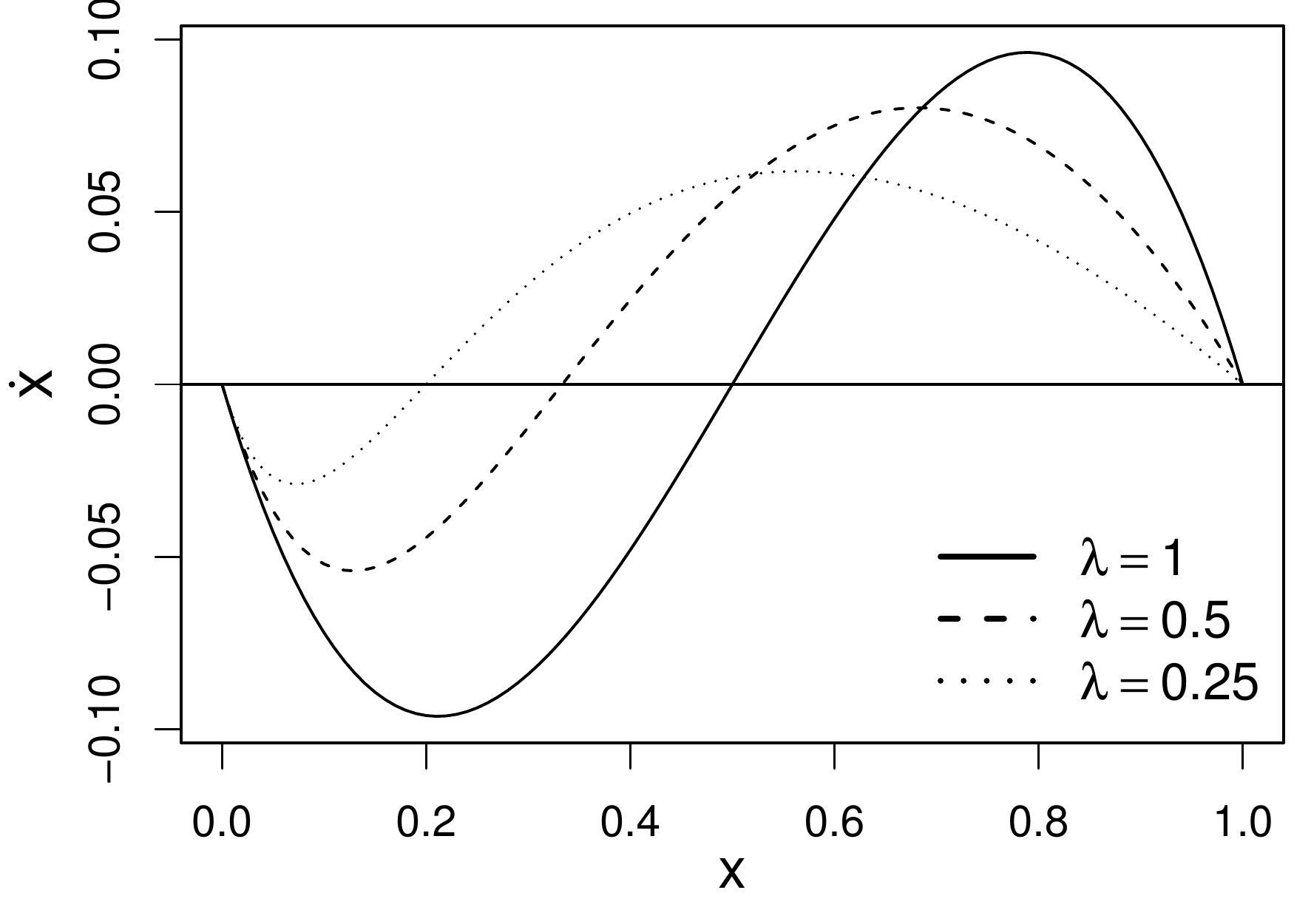}
\caption{Change of the number of voters favouring opinion \opi~dependent
on the actual fraction of voters with this opinion}
\label{figure:SM}
\end{figure}

Figure \ref{figure:SM} depicts $\dot{x}$ for $\lambda \in \{1,0.5,0.25\}$. The
zeros of $\dot{x}$, i.e., the stationary solutions of (\ref{equation:sdes})
are the (stable) consensus states [$x=0$] and [$x=1$] and the (unstable)
equilibrium point [$x=\lambda/(1+\lambda)$]. The latter marks the critical density that separates the flow to the consensus states.

To estimate the time until consensus for a finite number of voters we
rewrite equation (\ref{equation:sdes}) in the partial fraction expansion and integrate over
a suitable chosen interval. More precisely, we integrate from point $a_0=n/N$ to
a point $a_\infty$ sufficiently near the respective consensus state. The point $a_\infty$
corresponds to a state in which the system deviates from consensus only in a
single voter. Hence, we thereby get an estimation of the time
it takes the system to reach a state where only one last application of the
majority is needed to reach consensus. For any state
$a_0>\lambda/(1+\lambda)$ greater than the critical density the system finds
consensus on \opi, whereas for $a_0 < \lambda/(1+\lambda)$ the system will
develop consensus on \opii. Thus, the time to reach consensus in a system of $N$
voters that starts with $n$ voters for opinion \opi~can be approximated as
\begin{equation}
T^N_{n}\approx\int_{\frac{n}{N}}^{a_\infty} \left [
\frac{4}{a(\lambda+1)-\lambda}-\frac{1}{(a-1)\lambda}-\frac{1}{a} \right ] da
\end{equation}
with
\begin{equation*}
a_\infty=
\begin{cases} 0+ \frac{1}{N} & \text{if $\frac{n}{N}<
\frac{\lambda}{1+\lambda}$,}
\\
1-\frac{1}{N} &\text{if $\frac{n}{N} > \frac{\lambda}{1+\lambda}$.}
\end{cases}
\end{equation*}

\begin{figure}[h]
\centering
\includegraphics[width=0.45\textwidth]{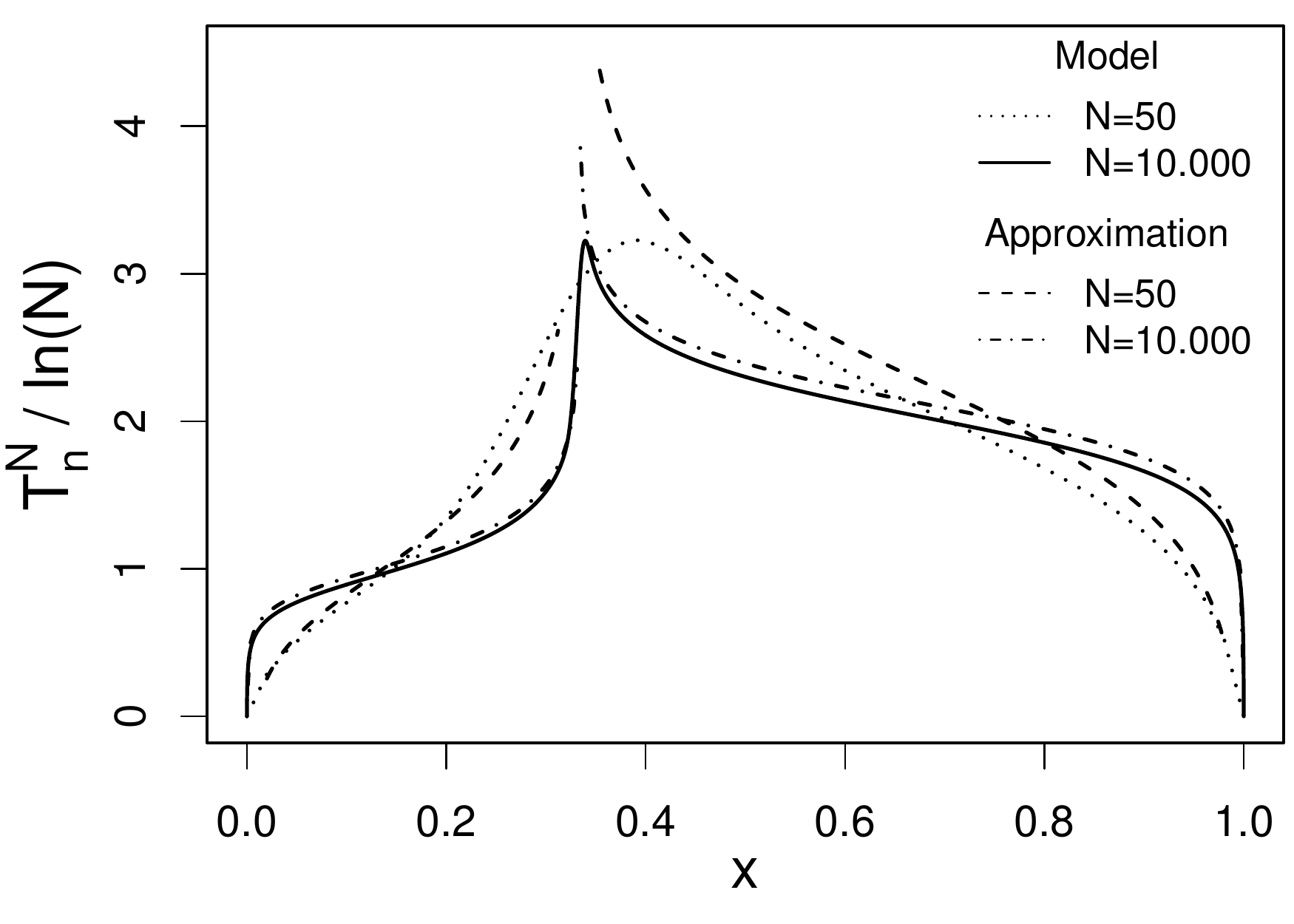}
\caption{Consensus time $T_n^{N}$ versus $x=n/N$ (initial fraction of voters
for opinion \opi) for different number of voters}
\label{figure:ttc}
\end{figure}

Figure \ref{figure:ttc} depicts results of this approximation together with
solutions of the model (\ref{eqn:9}). For few voters ($N=50$) the models differ
the most near the critical density. This is because systems that are initially
biased towards one opinion still can reach consensus on the other opinion. This
fact is taken into account by the model obtained from the master equation. The
approximation model, on the other hand, assumes that the critical density
determines the fate of the system. However, for larger $N$ the approximation
becomes more accurate (compare results for $N=10.000$ in Figure
\ref{figure:ttc}).

In the following we determine $T_{max}$, the maximal time until consensus is
reached for a given number of voters $N$. To estimate this time we integrate from a point
that deviates only in one voter from the critical initial density:

\begin{equation}
T_{max}\approx T^{N}_{\left [ \frac{\lambda}{1+\lambda}N+1 \right ] } \sim
\frac{5\lambda + 1}{\lambda(1+\lambda)}\ln N
\end{equation}

For $\lambda=1$ and large $N$ this result reduces to the asymptotic behaviour
$T_{max} \sim 3 \ln N$ that was also derived in \cite{CheRed05} for the standard
MR model. Moreover, as long as the latencies for the two opinions are comparable
the maximal consensus time grows asymptotically as $T_{max} \sim \ln N$. However,
if only very few voters for \opii~go into non-latent state the consensus time is
mainly determined by this flow rate. This is reflected by the fact that for very
long latencies $\lambda \ll 1$ the consensus time grows as $T_{max} \sim
1/\lambda$.

If we consider densities sufficiently far from the critical density, that is, if
$x-\lambda/(1+\lambda)$ becomes comparable to either $0$ or $x$ the consensus
time $T^N_{n}$ drops quickly (see Figure \ref{figure:ttc}). For the MR model
without latencies such a change in the amplitude in the consensus time to
$T^N_{n} \sim \ln N$ was also mentioned in \cite{CheRed05} and in
\cite{Krapivsky03}. However, in the case of differential latencies the drop of
the amplitude is not symmetrical. As already explained, at a certain point in the
evolution of the system the time until consensus is mainly determined by the rate
the voters that are in the minority leave the latent state. This is the reason
why for $x<\lambda/(1+\lambda)$ the consensus time drops to $T^N_{n} \sim \ln N$,
but for $x>\lambda/(1+\lambda)$ it drops to $T^N_{n} \sim 1/\lambda \ln N$.

\begin{center}
{\bf\small Conclusions}
\end{center}

The introduction of differential latencies in the MR model leads to a bias
towards consensus on the opinion that is associated with the shorter latency.
This effect increases with the number of voters $N$ and with the ratio
between the latency times $\lambda$. Moreover, the maximal time to
find consensus scales as $1/\lambda \ln N$.

These results particularly apply for systems of voters that start unbiased (i.e.,
with initial density $x=0.5$). Hence, from the point of view of the application
in a decision making method \cite{montes10}, this confirms two important
scalability results. First, the decision method improves when the
number of agents is increased and second, the time needed to find a decision is,
however, only marginally influenced by an increased number of agents.

\begin{center}
{\bf\small Acknowledgement}
\end{center}

I thank M. Montes de Oca, M. Birattari and Prof. M. Dorigo for their
helpful comments. This work was supported by a fellowship within the Postdoc-Programme of the
German Academic Exchange Service (DAAD).

\bibliographystyle{abbrv}

\bibliography{biblio.bib}

\end{document}